\documentclass[traditabstract]{aa} 

\usepackage{natbib,amsmath,amssymb}

\newcommand{\hi}{\mbox{H{\sc i}}}

\newcommand{\msol}{\rm M$_\odot$}

\newcommand{\kms}{km~s$^{-1}$} 
 
\usepackage{graphicx,amssymb,amsmath,natbib,hyperref,multirow}
\bibpunct{(}{)}{;}{a}{}{,} 
\usepackage{rotating}
\usepackage{txfonts}

\begin{document}

\title{The incorrect rotation curve of the Milky Way}
   \titlerunning{The incorrect rotation curve of the Milky Way}

 \author{Laurent Chemin\inst{1,2}
 \and Florent Renaud\inst{3}  \and Caroline Soubiran\inst{1,2}}

\institute{Univ. Bordeaux, LAB, UMR 5804, F-33270, Floirac, France 
\and CNRS, LAB, UMR 5804, F-33270, Floirac, France -  \email{astro.chemin@gmail.com}
\and  Department of Physics, University of Surrey, Guildford, GU2 7XH, UK}

   \date{Received \today; accepted XX}

  \abstract{In the fundamental quest of the rotation curve of 
  the Milky Way, the tangent-point method has long been the simplest way 
  to infer velocities for the inner  low-latitude regions of the Galactic disk from observations of the gas component. 
    In this article, we  test 
  the validity of the method on a realistic gas distribution and kinematics of the Milky Way, 
  using a numerical simulation of the Galaxy. 
  We show that the resulting velocity profile strongly deviates from the true rotation curve of the simulation because it 
  overstimates it in the central regions and  
  underestimates it around   the bar corotation. In addition, its shape is strongly dependent on the orientation of the stellar bar with respect to the observer. 
  The discrepancies are caused by the highly nonuniform nature of the azimuthal velocity field and by the 
   systematic selection by the tangent-point method  of high-velocity gas along the bar and spiral arms, 
   or low-velocity gas in less dense regions. The velocity profile  only agrees well with the rotation curve beyond corotation,  
  far from massive asymmetric structures. 
  Therefore the observed velocity profile  of the Milky Way inferred by the 
  tangent-point method is expected to be very close to the true Galactic rotation curve for $4.5 \protect\la R \le 8$ kpc. 
  The gaseous  curve is flat and consistent with rotation velocities of masers, 
  red clump, and red giants stars measured with VLBI astrometry and infrared spectroscopy for $R \ge 6$ kpc.
  Another consequence is that the Galactic velocity profile for $R< 4-4.5$ kpc is very likely flawed by the nonuniform azimuthal velocities 
  and does not represent the true Galactic rotation curve, but instead local motions.  
  The real shape of the innermost  rotation curve is probably  shallower than previously thought.   
   Using an incorrect rotation curve has a dramatic effect on the modeling of the mass distribution, 
   in particular for the bulge component, whose derived enclosed mass within the  central kpc and scale radius  
   are, respectively, twice and half of the actual values. We  therefore strongly argue against using 
   terminal velocities or the velocity curve  from the tangent-point method 
  to model the mass distribution of the Milky Way. The quest to determine the innermost rotation curve of the  Galaxy  remains open. }
   

   \maketitle

\section{Introduction}
\label{sec:intro}

The quest of the velocity field  of the Milky Way has long been  a very difficult 
task. The location of the solar system inside the  disk  makes it impossible to directly constrain  the 3D
  position and velocity phase space  for the whole   Galaxy. 
 Most surveys devoted to the chemistry, kinematics, and dynamics   of the Galactic disk(s) 
 are based on optical  spectroscopy, which is  not  appropriate  to probe the innermost low-latitude disk regions. 
   It can only be surveyed at infrared, mm-, and cm-wavelengths where  obscuration by dust is less problematic. 
  This is one of the objectives  of  the infrared Sloan Digital Sky Survey III's Apache Point Observatory Galactic Evolution Experiment \citep[APOGEE][]{eis11}. 
  For instance, \citet{bov12} have modeled the kinematics of 3365 red giants and red clump stars from the first year of APOGEE data and 
  found    a   rotation curve   consistent with a power-law model almost perfectly flat within  $4\la R \la 14$ kpc.  
   Line-of-sight velocities of stars that yield indirect estimates of distances  and rotation velocities 
   are not efficient in probing the   rotation curve in the innermost regions, however, which would be necessary to 
   establish accurate mass distribution models. 
  The best technique currently in use to do this is probably astrometry of methanol and water masers in high-mass star formation regions 
  with Very Long Baseline Interferometry   \citep[VLBI, e.g.,][]{bru09,rei09,asa10,sho11,hon12,ryg12,xu13,zan14}. 
  Preliminary results from these experiments combined within the Bar and Spiral Structure Legacy 
  (BeSSeL) Survey 
   have   for the first time directly revealed the  spiral structure of the Galaxy and 
   showed an apparently shallow inner rotation curve \citep{rei14}. 
  Although very promising,  it is too early for BeSSeL to entirely constrain the 
  innermost rotation curve at a statistically significant level,
however, since 
  the current number of reliable velocities remains small, with fewer than ten sources inside $R =4$ kpc. 

 Instead,  CO and \hi\   observations of the gaseous interstellar medium 
 have long been used as the reference to establish the inner rotation 
 curve of the Milky Way \citep{bur78,gun79,cle85,fic89,mcc07,lev08,sof09,mar12}.   
 Indirect estimates of rotation velocities and distances are obtained from the tangent-point (TP) method,   
which uses terminal line-of-sight velocitites  (see Sect.\ref{sec:tpmmw}). Such velocities and the rotation curve inferred from the TP method
 are the kinematical basis of many  Galactic mass models  
  \citep[e.g.,][]{mer92,deh98,kal03,fam05,sof09,sof13},   3D models of the inner neutral atomic and molecular gas 
  disks of the Galaxy \citep{nak03,nak06,kal07}, or   Galactic  population synthesis models \citep{rob03}. 
 
  However, this curve does not perfectly reflect the reality  
  because the gas distribution and kinematics are perturbed by 
  the Galactic bar. The effects of the bar on  
   gas terminal velocities and longitude-velocity diagrams in the central regions were first shown by \citet{lis80} and \citet{bli91}. 
  Numerical simulations of the Galactic interstellar medium 
  have shown the dependency of such  diagrams 
   on the position of the observer in the disk \citep{fux99,eng99,bis03,rod08}. 
 As a consequence, most of mass models  restricted the analysis  to  $|l| > 15-20\degr$ or $R>0.2 R_0$ ($l$ is the longitude, 
 $R$ is the Galactocentric radius and $R_0$ the Galactocentric radius of the Sun), 
 to minimize the perturbing effects from the bar at the center. 
 Although laudable, these arbitrary limits remain questionable because the bar and the spiral structure 
  are expected to  have a significant effect over more extended regions. 

 In this paper we   propose to test the capacity of the tangent-point method  in 
 predicting a reliable rotation curve using a  hydrodynamical simulation of a Milky Way-like galaxy \citep{ren13}.
 This simulation sets up a live model of the Galaxy in a self-consistent way, that is, without a 
 fixed pattern speed. The resulting structure and velocity field of the stellar and gaseous components
  are  far from being axisymmetric and circular. In particular, noncircular velocities as high as 200 
  \kms\ are found along the major axis of the bar.
 We investigate  the effect of the nonaxisymmetric  perturbations on 
 the shape and amplitude of the inferred curve and on mass models. We analyze the properties of tangent points  
 and constrain the radial range over which the tangent-point method provides a reliable estimate of the actual rotation curve. 
 The basics of the tangent-point method  are  
 described in Sect.~\ref{sec:tpmmw}, the confrontation to the gas simulation 
 and a discussion of the limits of the method is provided in Sects.~\ref{sec:tpmsimu} and~\ref{sec:limits}, 
 and the consequences for mass distribution modeling and for the Galaxy are detailed in Sect.~\ref{sec:discussion}.
  
 \begin{figure}[t]
 \begin{center}
  \includegraphics[width=0.9\columnwidth]{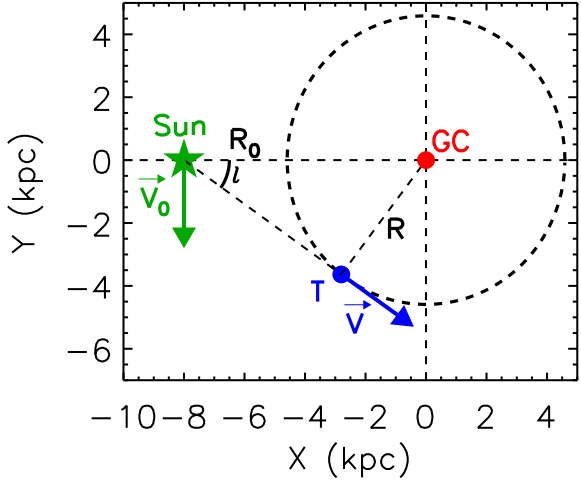}
   \caption{Geometry of the system: Sun, tangent-point (T), Galactic Center (GC) assumed by the tangent-point method. 
   The Sun is at radius $R_0$, T at radius $R$. The circular velocities are $v_0$ at $R_0$, 
   and $v$ at $R$, $l$ being the Galactic longitude. We choose here a view  similar  as in \citet{ren13} where 
   the Northern and Southern Galactic Poles are respectively in front of and behind the reader.}
 \label{fig:geom}
 \end{center}
 \end{figure}

\section{The tangent-point method applied to the Galaxy}
\label{sec:tpmmw}

For any given Galactocentric radius $R$ smaller than the solar radius, 
there are two line-of-sights at longitude $\pm l$ in the Galactic
plane that are tangent to the circle of radius $R$
 (Fig.~\ref{fig:geom}). The tangent-point  method stipulates that the  line-of-sight (l-o-s) velocity 
 $v_{T,\rm los}$ is an extremum at the position of the tangent
point $T$.
  The Galactocentric radius  and circular velocity profile $v(R)$ of such material at Galactic latitude $b=0\degr$ are 
   expressed as a function of  $l$ and the terminal velocity $v_{T,\rm los}$ by\begin{equation}
\label{eq:radiustpm}
R  =  R_0\sin(l)  \ ,
\end{equation}
\begin{equation}
\label{eq:vlostpm}
v(R)  =  v_0 \sin(l) + v_{T,\rm los} \ ,
\end{equation}
where $v_0$ is the circular velocity at the Galactocentric radius $R_0$.
 It is straightforward to 
estimate the rotation velocity profile  from Eq.~\ref{eq:vlostpm} with mm- or cm- observations  
by selecting the \hi\ or CO spectral component whose l-o-s velocity  $v_{T,\rm los}$ is lowest for $l < 0\degr$ and
highest  for $l > 0\degr$.

\begin{figure}[t]
 \begin{center}
  \includegraphics[width=0.9\columnwidth,trim=0 93 0 0,clip=true]{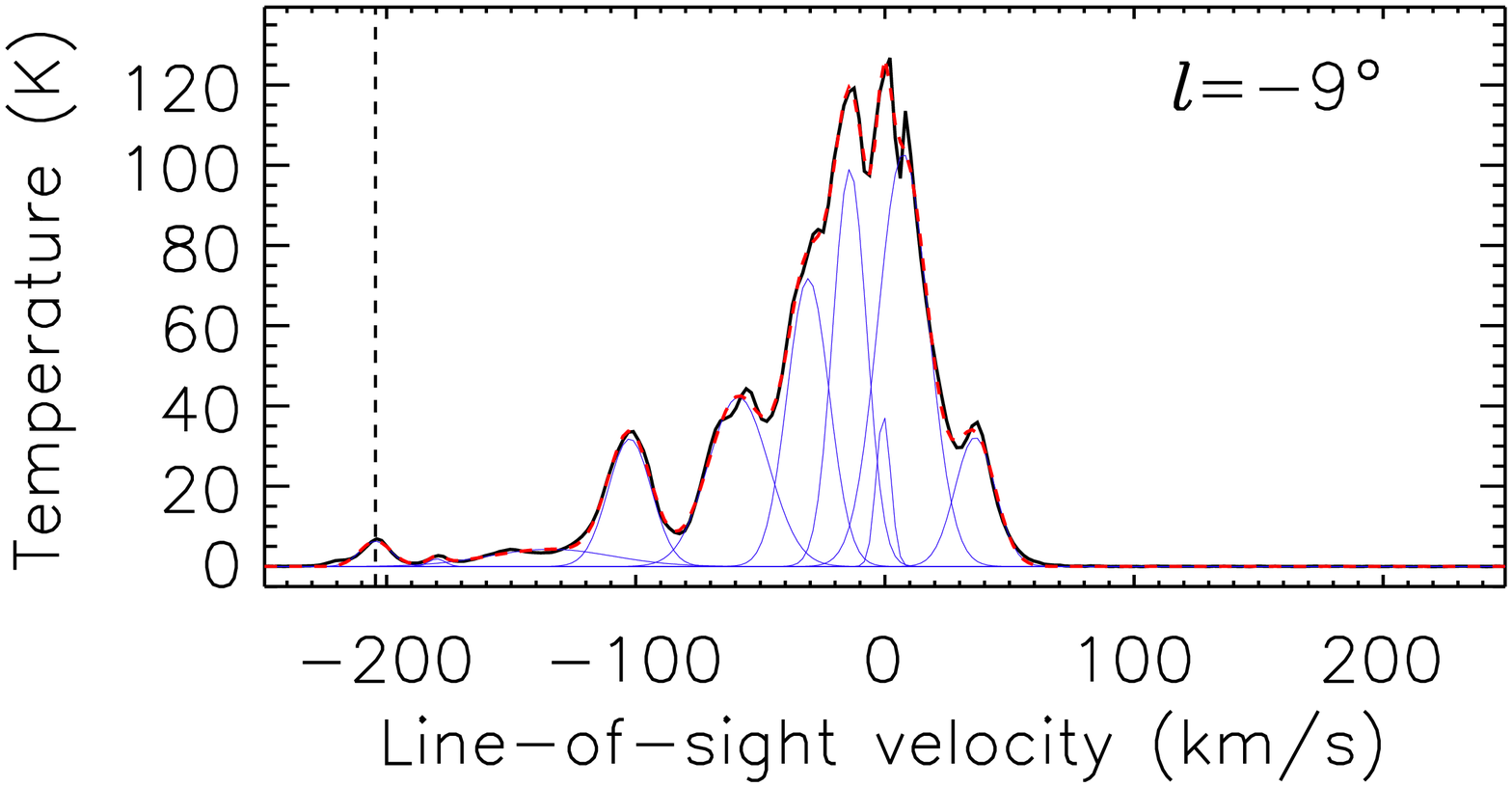}
  \includegraphics[width=0.9\columnwidth]{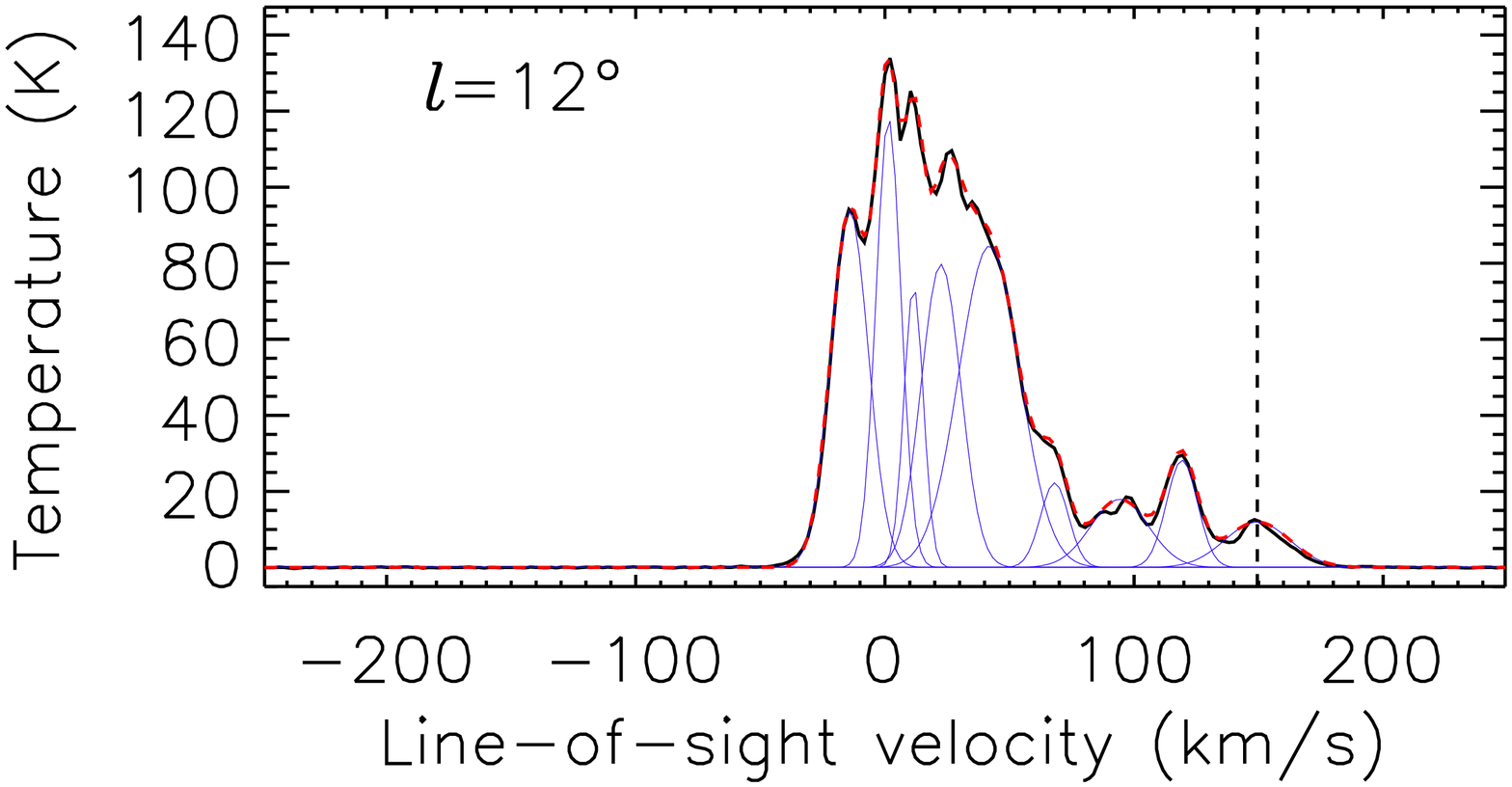}
   \caption{Examples of spectral decomposition of the \hi\ profiles at $l=-9\degr$ and $l=12\degr$. 
   The solid black curve shows the observed profile, which is decomposed using multiple Gaussian functions (in blue). 
   The reconstructed profile is shown as a dashed red curve.
   The vertical dashed line shows the terminal velocity of the  peak selected to derive the rotational velocity  with Eq.~\ref{eq:vlostpm}.}
 \label{fig:decomposition}
 \end{center}
 \end{figure}

We have  applied the TP method to the \hi\  Leiden/Argentine/Bonn  Survey datacube of the Milky Way \citep{kal05}. 
This survey yields the most sensitive all-sky  dataset of the neutral atomic  gas of the Galaxy
at angular and spectral samplings of 0.5\degr\ and 2 \kms. We used 
  $R_0=8$ kpc and $v_0=213$ \kms\ to facilitate comparison with the numerical 
simulation (Sect.\ref{sec:tpmsimu}). These are not the IAU values of 8.5 kpc and 220 \kms,  
but the choice of the fundamental parameters does not affect the result. \hi\ spectra were decomposed on a base of Gaussian functions, following the method used for the Andromeda galaxy
 \citep{che09}. In the example shown in Fig.~\ref{fig:decomposition},  the spectra
 were fitted with nine Gaussian profiles. The selected velocity extrema of  $\sim -205$ \kms\ at $l=-9\degr$  ($R=1.25$ kpc) and 
 $\sim 150$ \kms\ at $l=12\degr$ ($R=1.66$ kpc) correspond to   velocities of 238 \kms\ and 194 \kms, respectively. 
 
 \begin{figure}[h]
 \begin{center}
  \includegraphics[width=\columnwidth]{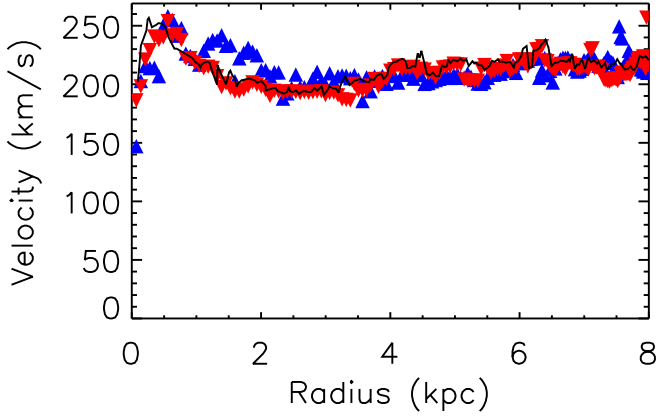}
   \caption{Velocity profile of the Milky Way  inferred with the tangent-point method applied to the \hi\ 
   LAB datacube of \citep{kal05}, assuming $R_0=8$ kpc and $v_0=213$ \kms. 
   Blue upward triangles are for $-90\degr \leq l  < 0\degr$,  red downward triangles  
   for  $0\degr \leq l \leq  90\degr$.  The solid line 
   is the composite curve from \cite{sof09} normalized at our value of $v_0$.}
 \label{fig:veltpmobs}
 \end{center}
 \end{figure}
   
By repeating this exercise for $-90\degr < l < 90\degr$, we obtained the 
velocity curves shown in Fig.~\ref{fig:veltpmobs}, which agree very well with 
 many other curves using the same method from CO
  or \hi\ data \citep[e.g.,][]{bur78,cle85,sof09,mar12}.    
 The steep profile in the center is clearly visible, as is the peak at $\sim 260$ \kms\ at 500 pc, 
 the smooth decrease to $\sim 200$ \kms,  the increase  beyond $R \sim 3.5$ kpc, and  flat profiles at large radii. 
 The axisymmetry between  the two quadrants is quite remarkable for $R > 3.5$ kpc.  
 The disk at $l<0\degr$ appears to rotate faster 
 than at positive longitudes by $\sim 17$ \kms\ on average 
for $1 < R < 3.5$ kpc. Note in particular the bump at $R\sim$ 1.5 kpc for $l<0\degr$ that is not measured for $l>0\degr$.
It has often been argued that   differences in the central regions  
 are signatures of perturbations, like the bar and/or the spiral structure. 
 This is a possible explanation, but we show in Sect. \ref{sec:results} that a velocity difference
 is not the best criterion to detect signatures of  perturbations. 
 Notice finally that the composite curve of \cite{sof09} is more consistent 
  with our velocities for $l>0\degr$ than for $l<0\degr$.  
 
\section{The tangent-point method to the test of numerical simulations}
\label{sec:tpmsimu}
\subsection{Mock   gas datacubes}
To quantify to which extent the TP method gives a reasonable rotation curve, 
we applied it to a simulation of the Milky Way, for which the rotation curve can be obtained directly 
from the velocity field. We used the simulation of \citet{ren13}, which reproduces  
the main structures (bar, spirals) of the  Galaxy both in terms of morphology and kinematics. 
 We produced mock observations by computing longitude, latitude, and line-of-sight velocity datacubes for an arbitrary position of the 
   Sun along the $R_0=8$ kpc circle. Varying the position of the Sun allows us to monitor the effects of the
 bar orientation  on the results from the TP method.  
 In our fiducial case, the bar major axis yields
  an orientation of 23\degr\ with respect to the direction of the Galactic center. 
   This angle  setup matches the real Galaxy best \citep{ren13}.  
     The covered range of bar orientations is 
   0\degr\ (Sun aligned with the bar) to 90 \degr\ (Sun - Galactic center axis perpendicular to the bar major axis). 
      It allows us to test, for instance, whether the bar viewing angle of $\sim$45\degr\ 
      inferred from some IR observations \citep{ham00,ben05} would yield results different from  our fiducial case. 
The velocities are computed from the gaseous (atomic and molecular) component of the model. 
 The spatial and spectral samplings of these mock datacubes are 2\degr\ for $(l,b)$ and 3 \kms\ for l-o-s velocity. These samplings are
 smaller than for the \hi\ observation,  implying smoother velocity profiles,  but this does not affect the  result of this work.
A   datacube was also generated using an earlier epoch of the simulation. It corresponds to an axisymmetric disk, that is,  
a reference case before the formation of the bar and the spiral arms. 
   
 \begin{figure}[t]
 \begin{center}
  \includegraphics[width=\columnwidth]{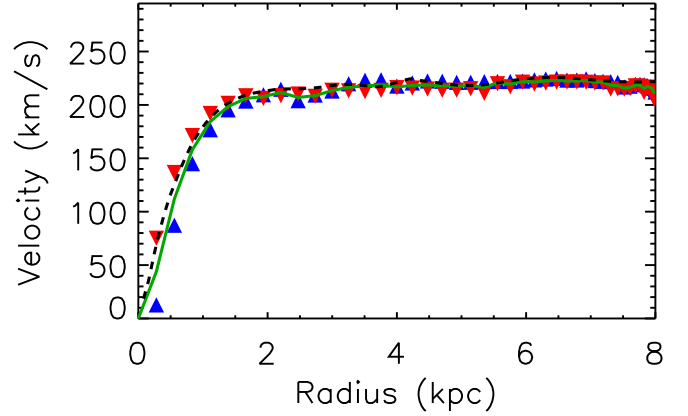}
   \caption{Rotation velocity profile of the simulated disk  inferred by the tangent-point method for the 
   axisymmetric  disk (before the formation of the bar and spiral arms). 
    Colored symbols are the same as in Fig.~\ref{fig:veltpmobs}. The solid line is the average curve from the two halves. 
    The dashed line is the true rotation curve of the simulated disk.}
 \label{fig:veltpm1}
 \end{center}
 \end{figure}

\subsection{Velocity profiles}
\label{sec:results}
Figure~\ref{fig:veltpm1}  shows the resulting velocity profile for the reference datacube of the axisymmetric
 disk. The 
velocity increases within the inner 1.5 kpc and  then remains constant with radius. 
The velocity curve averaged over positive and negative longitudes is exactly the one inferred from averaging the 
azimuthal velocity field  of gas cells from the simulation at every radii. 
This implies that there are no systematics in our approach and 
that the TP method can yield a velocity profile  that  excellently agrees with the true rotation curve  
 if the disk remains nearly axisymmetric.  
 \begin{figure*}[ht]
 \begin{center}
  \includegraphics[width=0.4\textwidth,trim=0 71 0 0,clip=true]{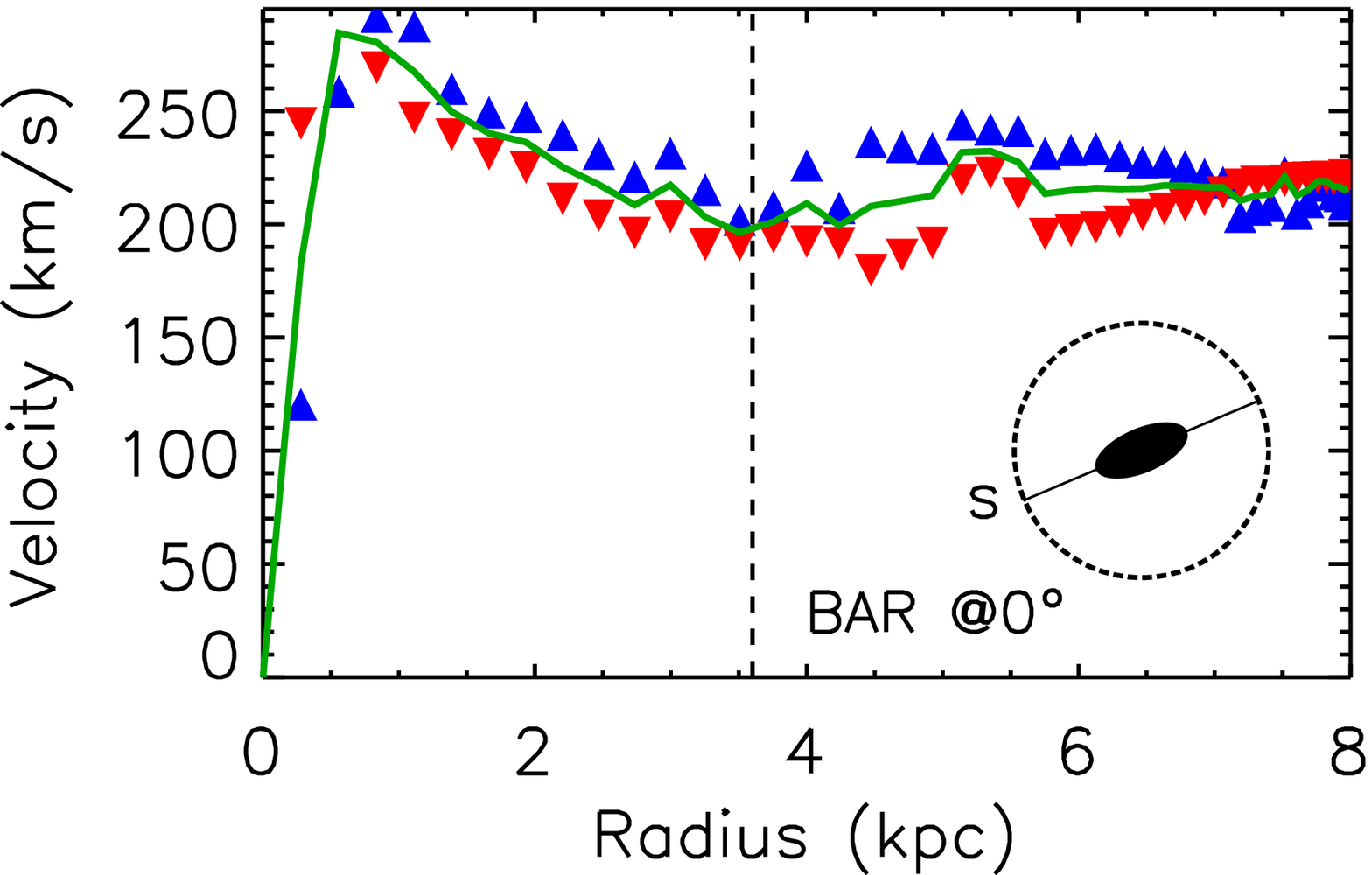}\includegraphics[width=0.4\textwidth,trim=0 71 0 0,clip=true]{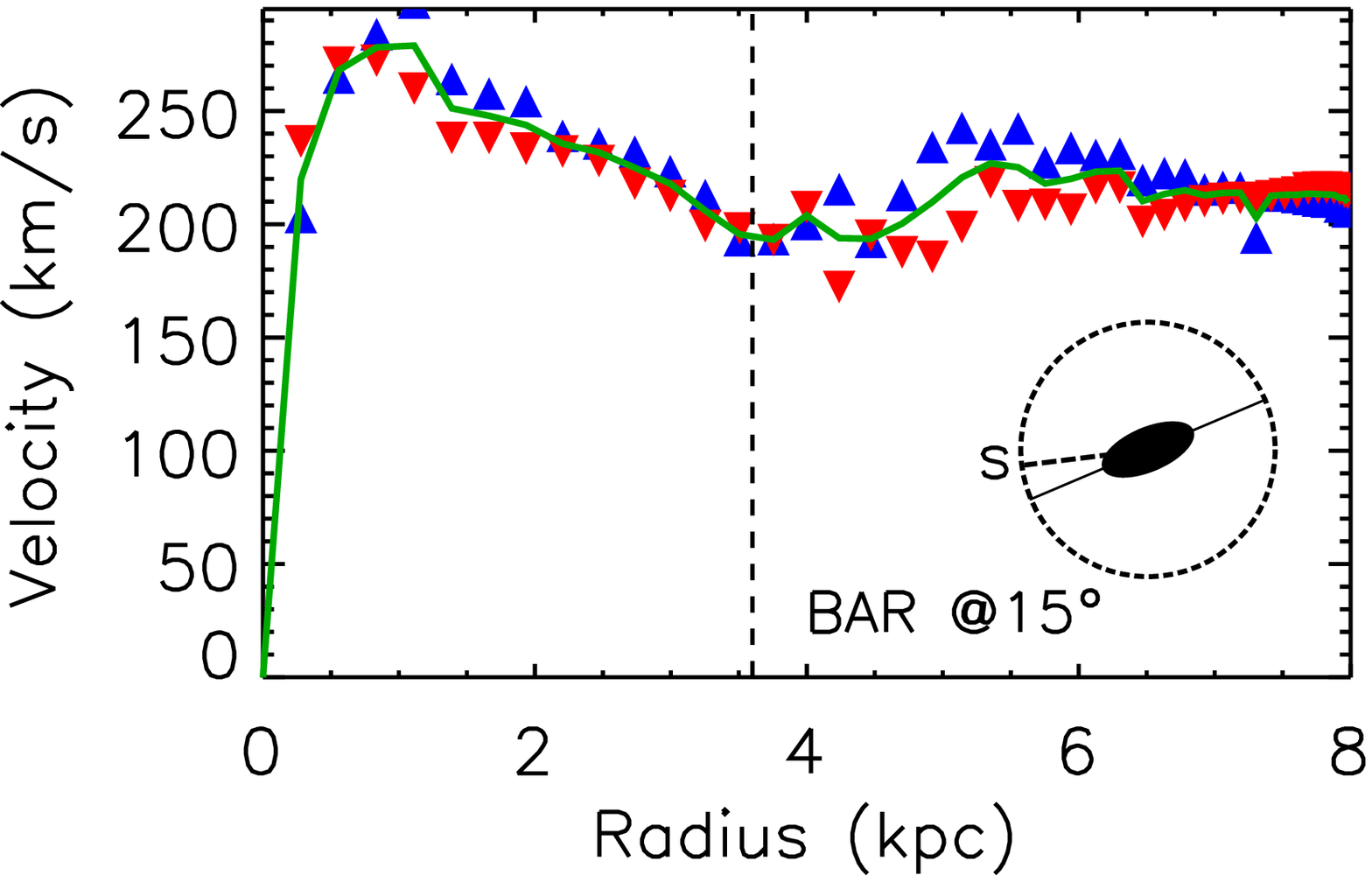}
  \includegraphics[width=0.4\textwidth,trim=0 71 0 0,clip=true]{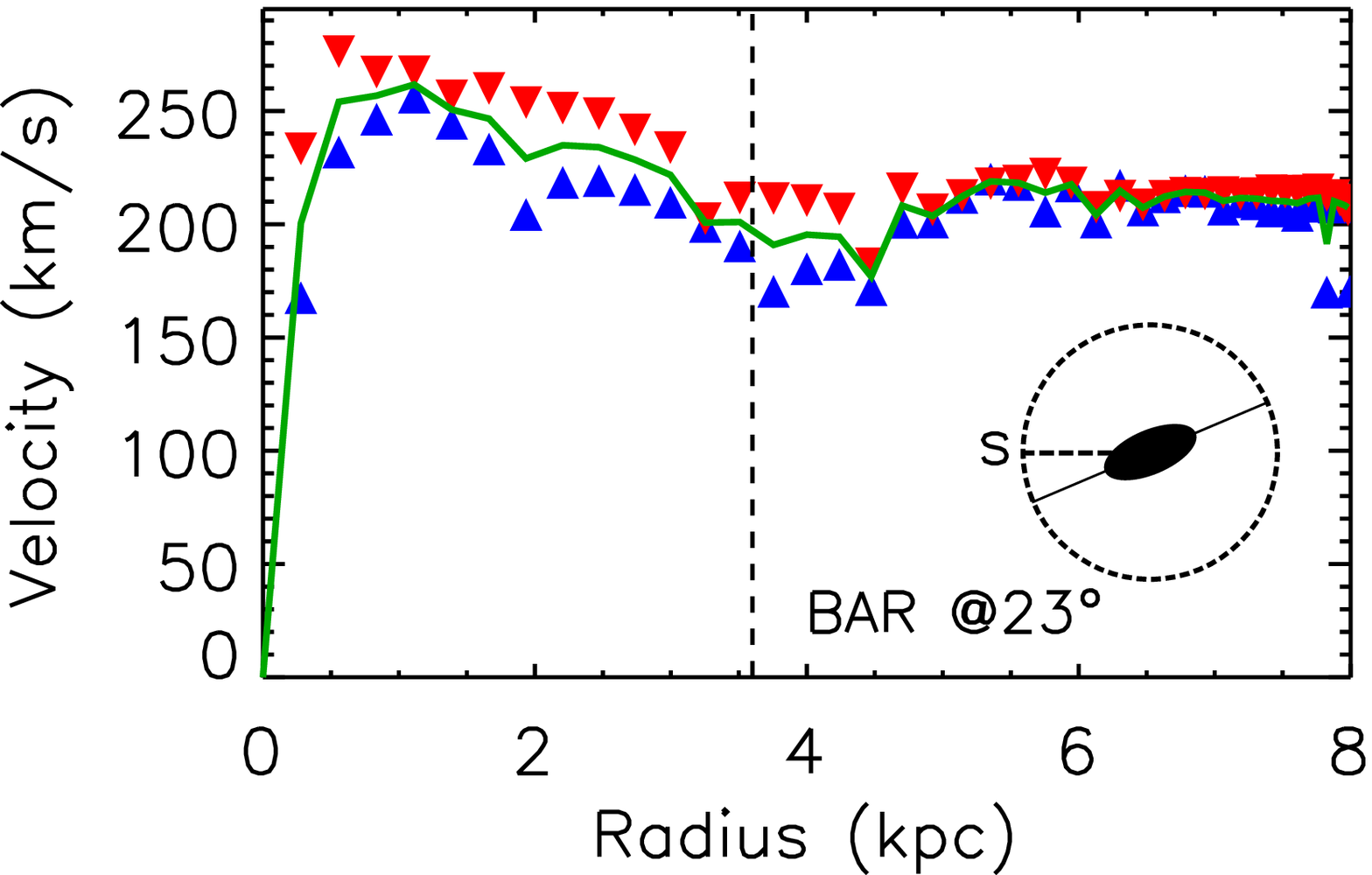}\includegraphics[width=0.4\textwidth,trim=0 71 0 0,clip=true]{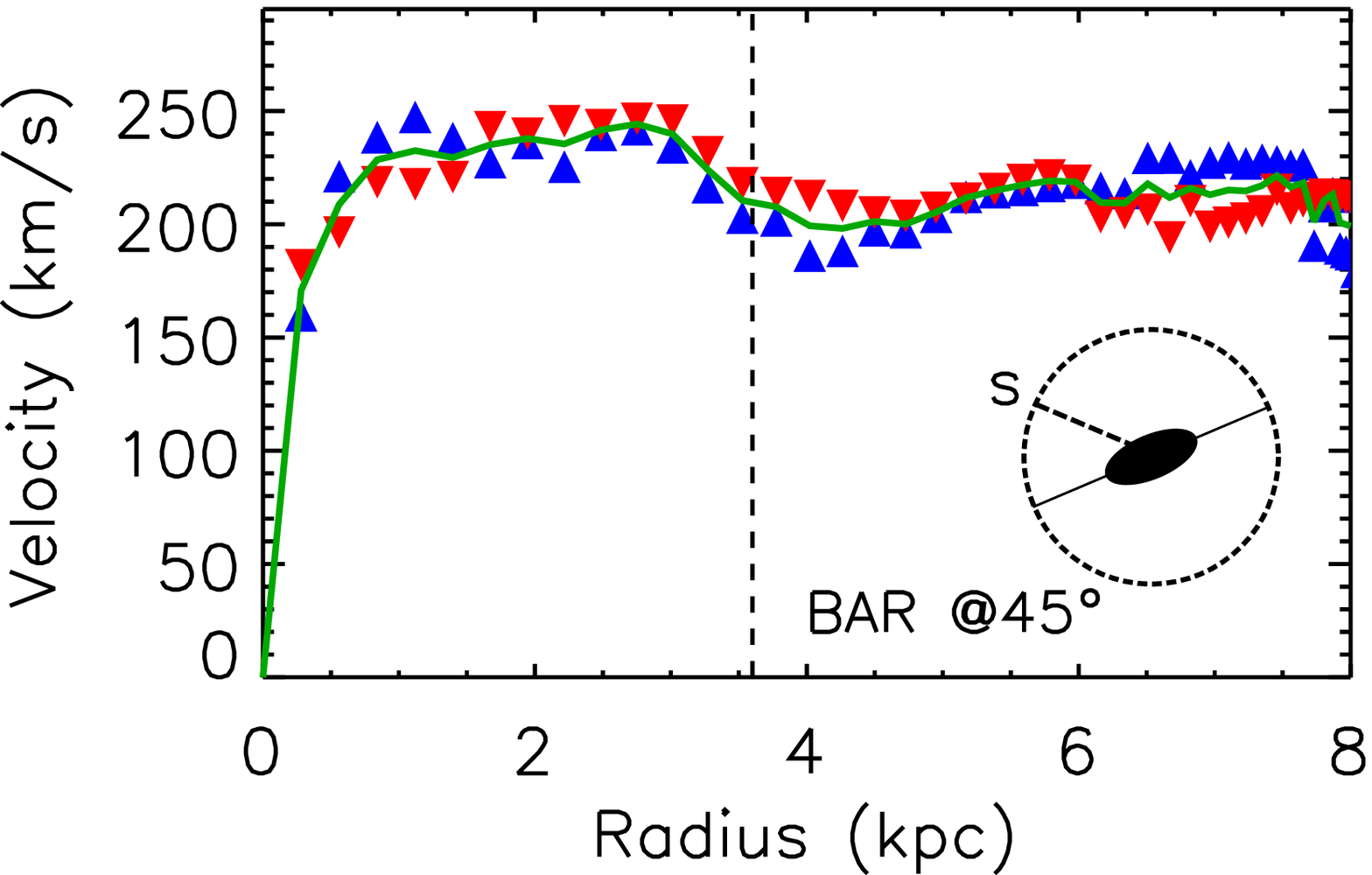}
  \includegraphics[width=0.4\textwidth,trim=0 0 0 0,clip=true]{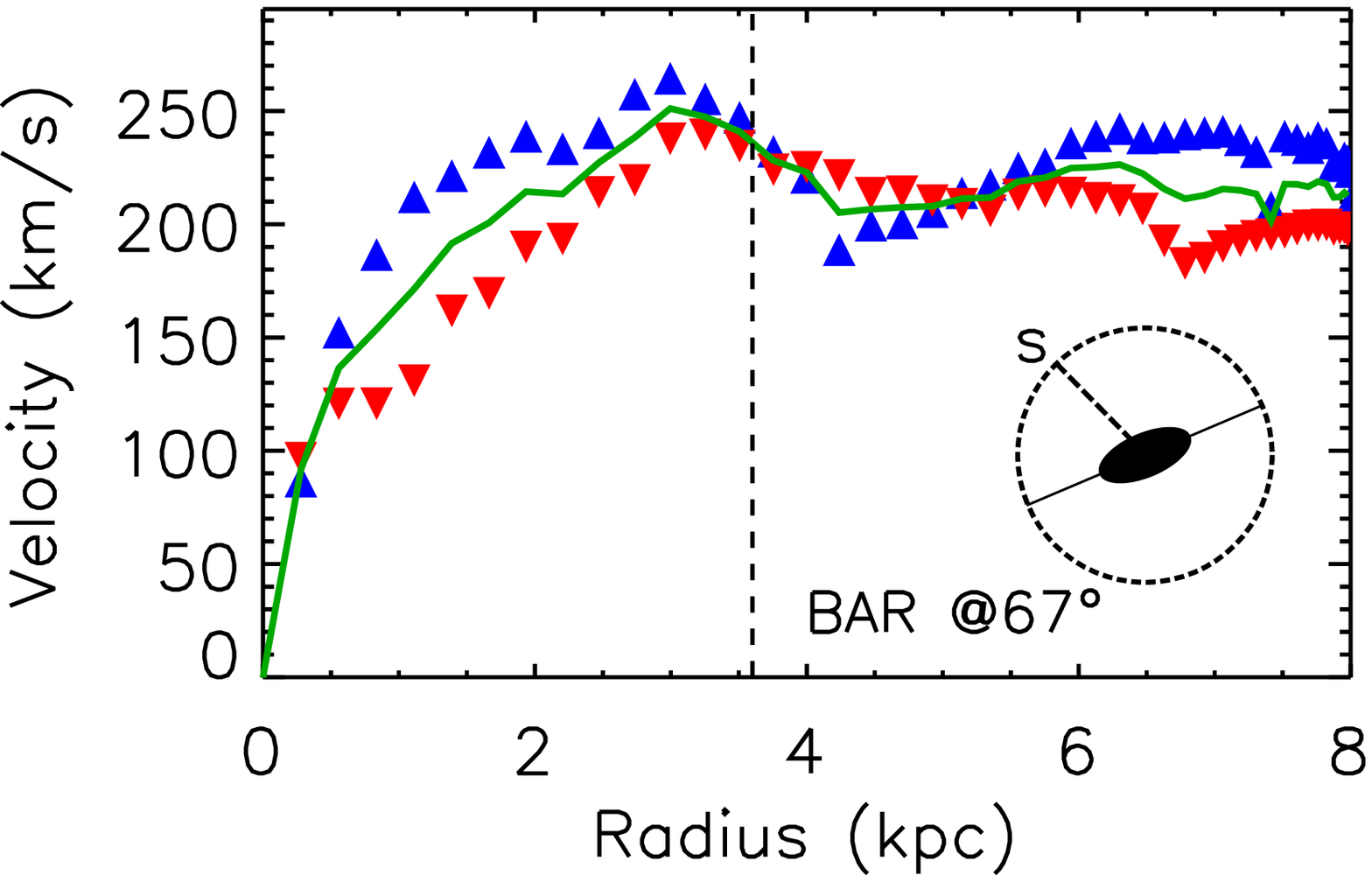}\includegraphics[width=0.4\textwidth,trim=0 0 0 0,clip=true]{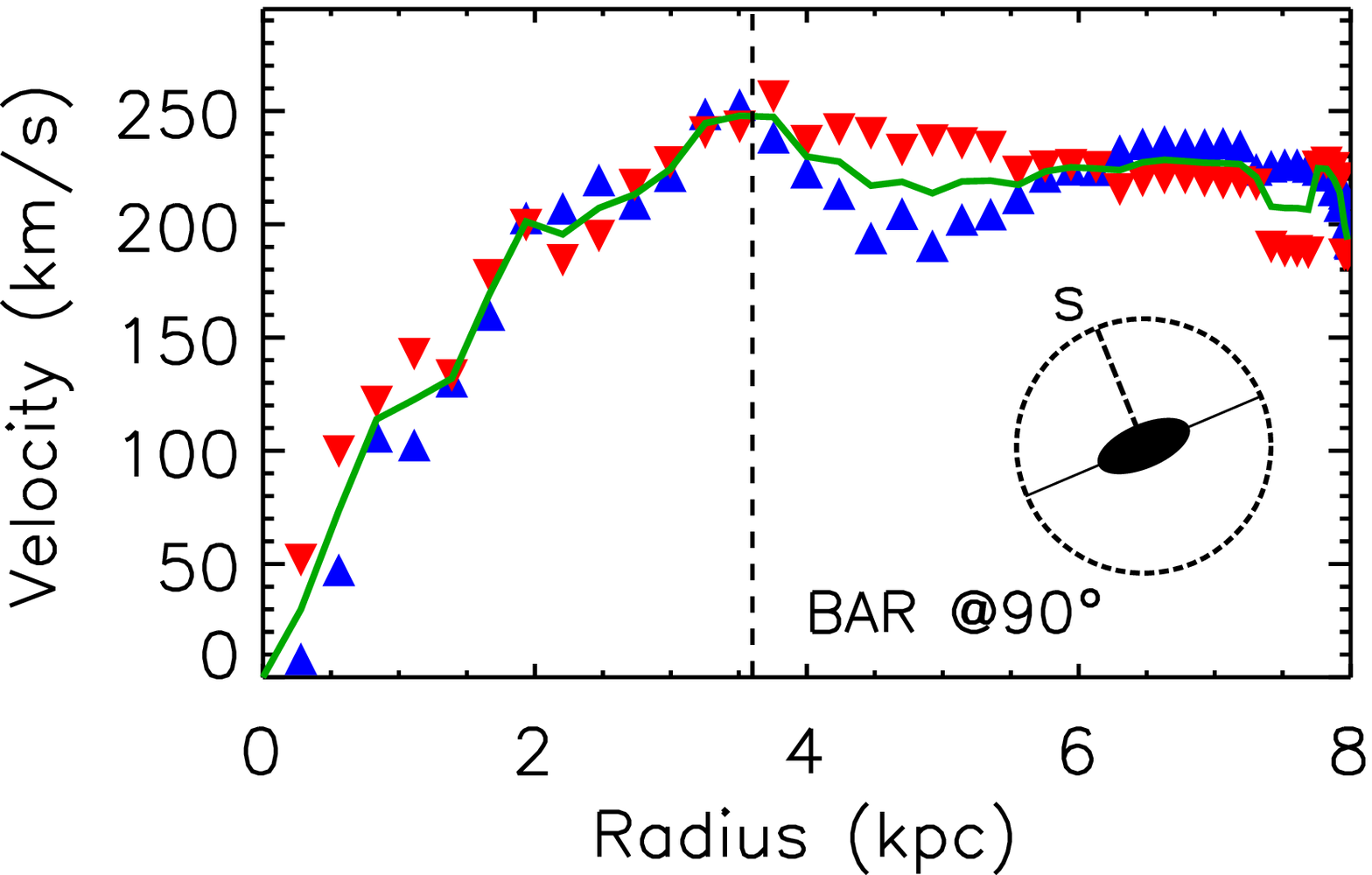}
   \caption{Rotation velocity profile of the simulated disk for several viewing angles of the bar. 
   The vertical dashed line is the radius of the bar corotation, as computed in \cite{ren13}.}
 \label{fig:tpmorientbar}
 \end{center}
 \end{figure*}

Using the snapshot that best matches the observations, that is, after the formation of the bar and the spiral arms, 
Fig.~\ref{fig:tpmorientbar} shows 
the  velocity profiles for a selection of several bar viewing angles. 
The shape and maximum amplitude of the velocity profile  strongly depend  on the bar viewing angle. 
The more aligned the reference point  with the bar major axis, the higher
 the velocity peak at small radius. Moreover, the slope of the  
velocity profile in the central region decreases with increasing viewing angle. 
Configurations with  angles between 0\degr\ and 23\degr\ all show a decrease of velocity from $R=0.5-1$ kpc 
 to $R \sim 4$ kpc, while the velocity  increases up to $R=3-3.5$ kpc for larger viewing angles. The velocity peak is thus offset for large viewing angles.
With the exception of the profile for  an angle  of 90\degr\ that continuously 
overestimates the true values beyond 3.5 kpc, 
 most  configurations yield velocities that agree fairly well with the correct rotation curve  
 for $R > 4.5$ kpc. In addition, the velocity is constant beyond 5.5 kpc.   
A common feature to most velocity profiles is the  
 inflection of the curve past the bar corotation ($R_c =3.6$ kpc).  
 
The velocity difference between the two quadrants strongly varies with the bar viewing angle. 
For instance,  the two quadrants yield very similar velocity curves at every radius for 
configurations of 15\degr\ or 45\degr, while larger differences are found at 0\degr\ and 67\degr.
The asymmetry is also stronger for $R< 4.5$ kpc   at 23\degr\ and 67\degr. 
The bar and spiral arms cannot be detected when viewed at 15\degr\ and 45\degr, 
with the assumption that  differences between the two quadrants trace nonaxisymmetric structures.
Similarly, it is difficult to reconcile the asymmetry
 at 0\degr\ and 67\degr\ at large radii where the  perturbations are weaker 
 with the axisymmetry at similar radii for all other configurations. 
Differences between the two quadrants thus cannot be the sole criterion to attest or reject 
the presence of nonaxisymmetric structures like the bar, or assess its strength.

\begin{figure}[h!]
 \begin{center}
  \includegraphics[width=0.9\columnwidth]{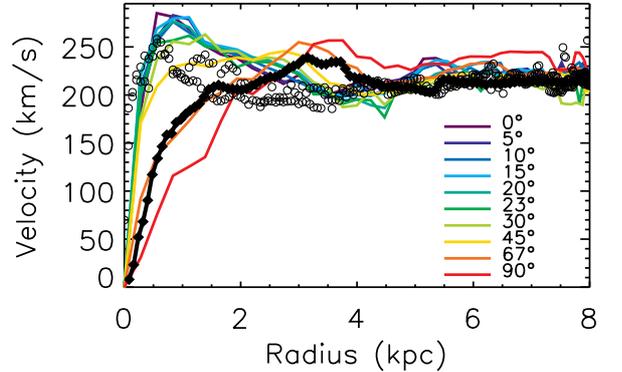}
   \caption{Comparison of the true rotation curve of the simulated disk (filled symbols) with the  velocity profiles inferred by the tangent-point method 
   (colored lines). The viewing angle of the bar  is indicated for each velocity profile. The open circles 
   represent the velocities from the \hi\ observations of the Milky Way.}
 \label{fig:comprcveltpm}
 \end{center}
 \end{figure}

The comparison between the  rotation curve of the mock Galactic disk and the  velocity profiles from the TP method 
all given at $(R_0,v_0)$=(8 kpc, 213 \kms)  is very instructive (Fig.~\ref{fig:comprcveltpm}). 
The tangent-point method  rarely gives a velocity profile   that agrees 
with the rotation curve for $R< 4.5$ kpc.  The profiles for  bar viewing angles $\la 45\degr$ 
overestimate the rotation curve inside $R=2.5-3$ kpc 
and then underestimate it past this radius. Almost all profiles are consistent with the true rotation  beyond $R=4.5$ kpc. 
The extreme viewing angle of 90\degr\ is the only model that uniformly
 overstimates the rotation curve at these radii, while it understimates it in the inner regions.
 To illustrate the  significant discrepancy at small radius, the difference of velocities for the fiducial 
 bar orientation of $23\degr$  is 100\%  of the true velocity at $R=0.5$ kpc, $\sim$50\% at $R=1$ kpc, 
 and 18\%  at $R = 3.7$ kpc, close to the bar corotation. 
 
\begin{figure*}[t]
\begin{center}
 \includegraphics[width=\columnwidth]{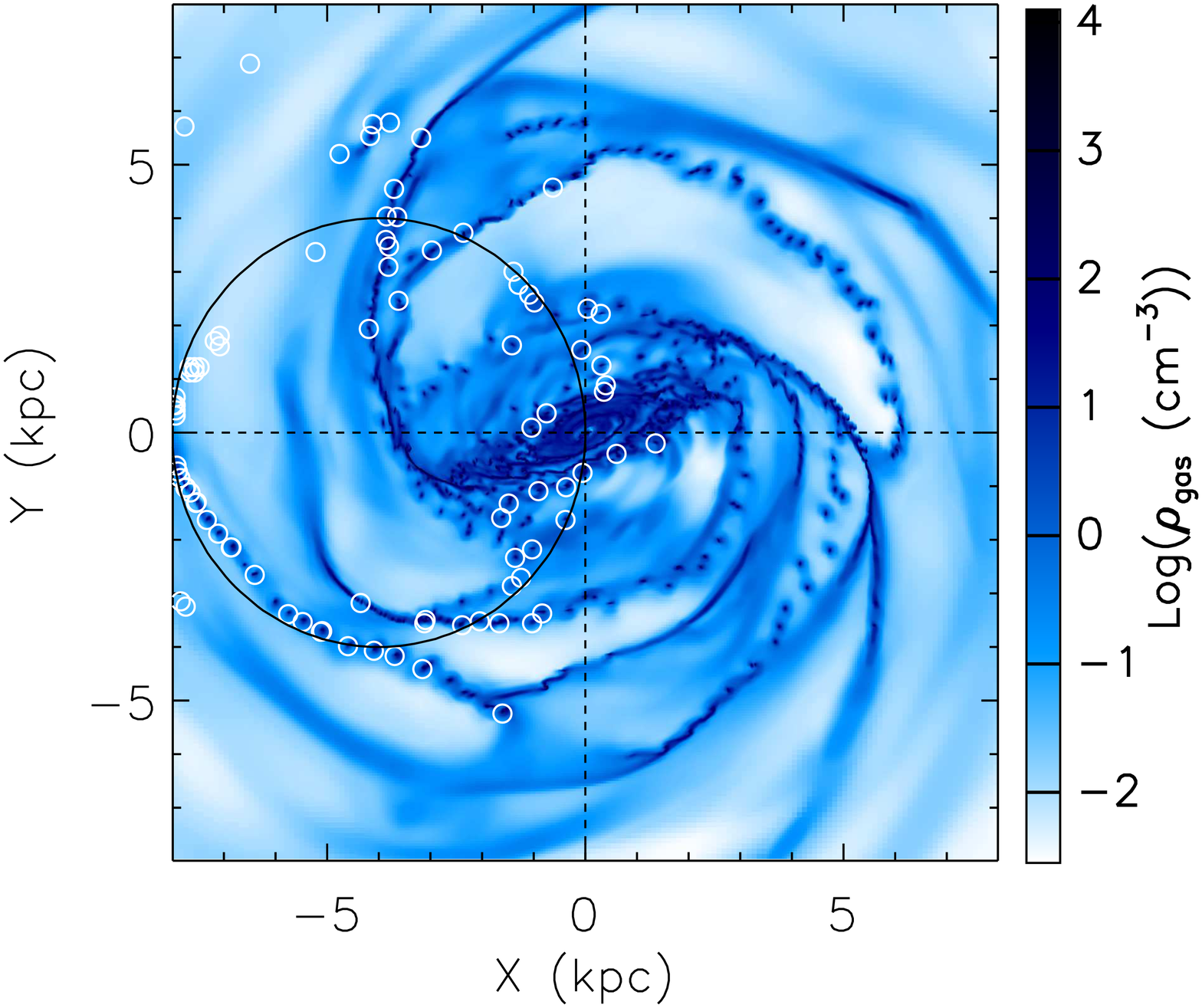}\includegraphics[width=\columnwidth]{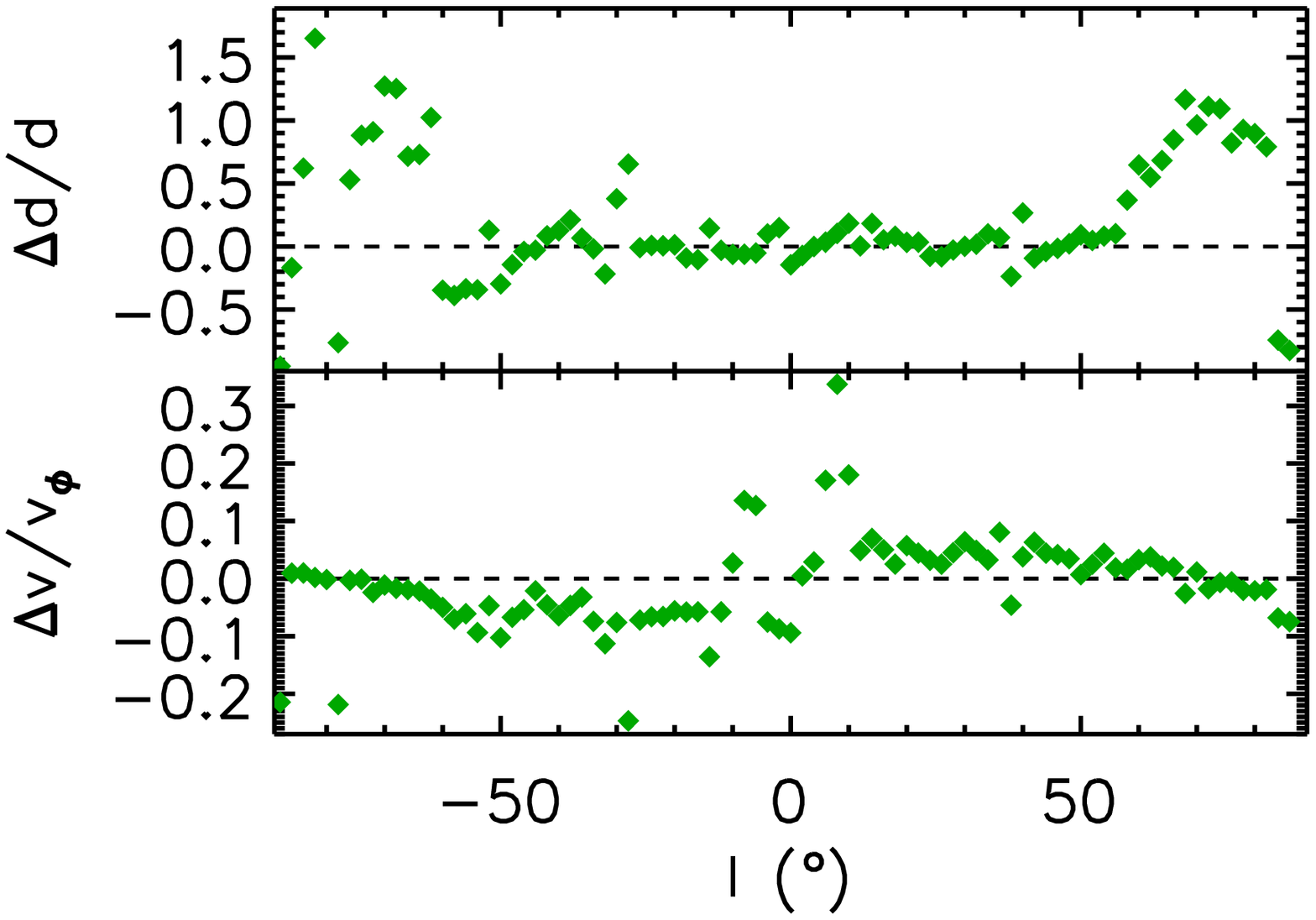}
\caption{\textbf{Left:} Gas density map showing the locations of terminal l-o-s velocity points (white circles). 
   The black circle indicates the positions of the tangent points. The reference  position is $(x,y)=(-8,0)$ kpc.
\textbf{Right:} Comparison of the distance (top) and velocity (bottom) of the tangent points with 
the true distance and velocity of the terminal velocity points. $\Delta d=d_{\rm T}-d$ 
where $d_{\rm T}=(R_0^2-R^2)^{1/2}$ is
  the  distance to   tangent points and $d$  the true distance  to  terminal velocity points. 
$\Delta v=v(R)-v_\phi$ , where  $v(R)$ is from Eq.~\ref{eq:vlostpm} 
and $v_\phi$ is the true azimuthal velocity of   terminal velocity points.}
 \label{fig:wheretheterminalvelocitypointsare}
 \end{center} 
 \end{figure*}
 
 Figure~\ref{fig:comprcveltpm} also shows the comparison with the observational \hi\ data. 
 The agreement is very good at $R > 4-4.5$ kpc for most of bar orientations. 
Bar viewing angles of $23\degr$ and $30\degr$ are the models that best 
reproduce the central slope and the amplitude of the peak at $R=0.5$ kpc. Smaller angles overestimate the peak, while larger angles 
underestimate it. Angles of 67\degr\ and 90\degr\ are the least compatible models with the observed velocities.
On average, the shape and amplitude of velocities in the fourth quadrant ($l<0\degr$, upper part of the observational points) 
are more consistent with the models than in the first quadrant (lower part of the points). 

\section{Limits of the tangent-point method}
\label{sec:limits}
\subsection{Properties of tangent points and terminal velocity points}
Although the tangent-point method is 
appropriate for the early stage snapshot (Fig.~\ref{fig:veltpm1}), 
its validity is clearly questionable for the asymmetric disk. To address this problem, 
we  analyzed the characteristics of the tangent points
 and  terminal velocity points.  We based the analysis only on the fiducial case (bar at 23\degr) because it represents the real Galaxy best \citep{ren13}. 

We first address the question whether tangent points are actual terminal velocity points. 
This goal can be achieved by comparing the tangent points with the  positions and azimuthal velocities
 of the terminal velocity points.  We find the gas at the origin of the terminal velocities by searching 
 the clouds whose   l-o-s velocity is within $\pm$5 \kms\ from the terminal velocity. This velocity range   corresponds to 
 a window of $5\sigma$, where $\sigma$ is the spectral sampling of the mock datacube. 
 Since more than one cloud may contribute to shape the emission profile around the terminal velocity,   
  we expect to find a few lines of sight that harbor several clouds within the search window. In this case, 
 we kept the cloud with the brightest\footnote{We roughly estimate the brightness for the simulation as the gas density 
 divided by the square of the distance to the Sun.} contribution to the emission profile because it represents that population of clouds best. Note that a wider search window  only marginally changes the identified clouds.    

 Figure~\ref{fig:wheretheterminalvelocitypointsare} shows the corresponding locations of the terminal velocity points. 
 These points are almost exclusively clouds lying along the spiral arms and the bar. 
 A few positions are not associated with a specific cloud, but 
 correspond to less dense gas in   inter-arm regions (e.g., $x,y
  \sim -1.3,2.5$), or along
   a more diffuse spiral structure (e.g., $x<-6$, $y >0$). 
    However, it is striking  that    terminal velocity points that coincide perfectly
     with the tangent points whose locations are delineated by a circle are rare. 
     In particular, the pitch angle of the spiral arms passing at the  Sun position
      causes the tangent-point method to systematically overestimate the true distance at high longitudes.
     
      Figure~\ref{fig:wheretheterminalvelocitypointsare}    shows the 
     errors on the  true distance  to the terminal velocity points ($d$) when assuming
       that the distance is that of the tangent points   $d_{\rm T}=(R_0^2-R^2)^{1/2}$ 
       (see Fig.~\ref{fig:geom}).
     On  average, the difference in distance is 13\%  for $-55\degr <l < 0\degr$ and
     8\%  for $0\degr < l < 55\degr$.  That ratio increases at $|l| > 55\degr$, where it is $\sim 86\%$ on average. 
     Therefore, the closer to the Sun the tangent point, the less accurate the distance $d_{\rm T}$.
  Figure~\ref{fig:wheretheterminalvelocitypointsare} also shows the difference between the velocity of the tangent point as given in Eq.~\ref{eq:vlostpm} and 
the local azimuthal velocity\textup{ \textup{\textup{\textit{at the position of the terminal velocity point}}}.}  The tangent-point velocity underestimates (overstimates) the  azimuthal velocity at negative (positive) longitudes. The difference increases toward low longitudes. 
On average, a difference of 7\% is measured for $-50\degr < l < -10\degr$ relative to the local azimuthal velocity; this is
4\% for $10\degr<l<50\degr$. The error remains relatively constant within this longitude range. 
It is negligible at high (absolute) longitudes and can be larger than 15\% in the inner $|l| < 10\degr$. 
We conclude that the tangent points and terminal velocity points are thus quite similar for intermediate 
longitudes, but 
 are certainly not the same at the lowest and highest longitudes, or at
 particular outlier directions.  The effect of the significant difference of position relative to the true position 
 at large $l$ is negligible for velocities, however, since the rotation velocity does not 
 vary significantly at large radii. 

Similarly, we analyzed the differences between the tangent-point velocity 
and the local azimuthal velocity \textup{\textit{at the position of the tangent-points}}. Here we again find a strong dependency on longitude, 
but the scatter is about five times larger than what is shown in Fig.~\ref{fig:wheretheterminalvelocitypointsare}  
for terminal velocity points.   In other words, a tangent-point velocity is surprisingly less consistent with its local azimuthal velocity 
than with the azimuthal velocity at the position of the terminal velocity point. 

 Interestingly, we note that velocity profiles based either on the local azimuthal velocity (and position) at
the terminal velocity points or on the local azimuthal velocity at the tangent-points 
 differ significantly from the rotation curve  for $R< 5$ kpc. Their shape remains similar to $v(R)$ of Figs.~\ref{fig:tpmorientbar} and
 ~\ref{fig:comprcveltpm}, again with velocities  overestimated in the inner regions, 
 and is underestimated around the bar corotation.
 
The distance errors have previously been estimated by \cite{gom06}, but from more idealized 
gas flow simulations than the one we used, for a disk perturbed by a $m=2$ spiral  within 
the fixed potential of \cite{deh98}. They have found mean 
errors lower than 0.5 kpc, and larger errors in the spiral arms. Our result is thus quite consistent with their estimate, on average.  

\subsection{Origin of the discrepancy with the rotation curve}
We have shown that  Eqs.~\ref{eq:radiustpm} and~\ref{eq:vlostpm} are rough  
  -- and sometimes incorrect -- approximations of the local positions and azimuthal velocities of terminal velocity points  
and are even poorer estimates of the local azimuthal velocities at the coordinates of tangent points. 
However, the position and velocity errors are not issues as such. Indeed, it remains difficult to recover
 the  shape of the rotation curve of the mock disk, 
irrespective of the nature of the points or velocities chosen to calculate the velocity profile.  
At the same time, we have  shown that the  bar orientation with respect to an observer strongly affects the 
shape of the inferred velocity profile. 
Ignoring the effect of the bar therefore renders the TP method useless in regions where the bar dynamics dominates the motions,  
and one may thus accordingly  question its validity  in estimating the true rotation curve of a barred spiral  galaxy like the Milky Way. 

 \begin{figure}[t!]
\begin{center}
\includegraphics[width=0.9\columnwidth]{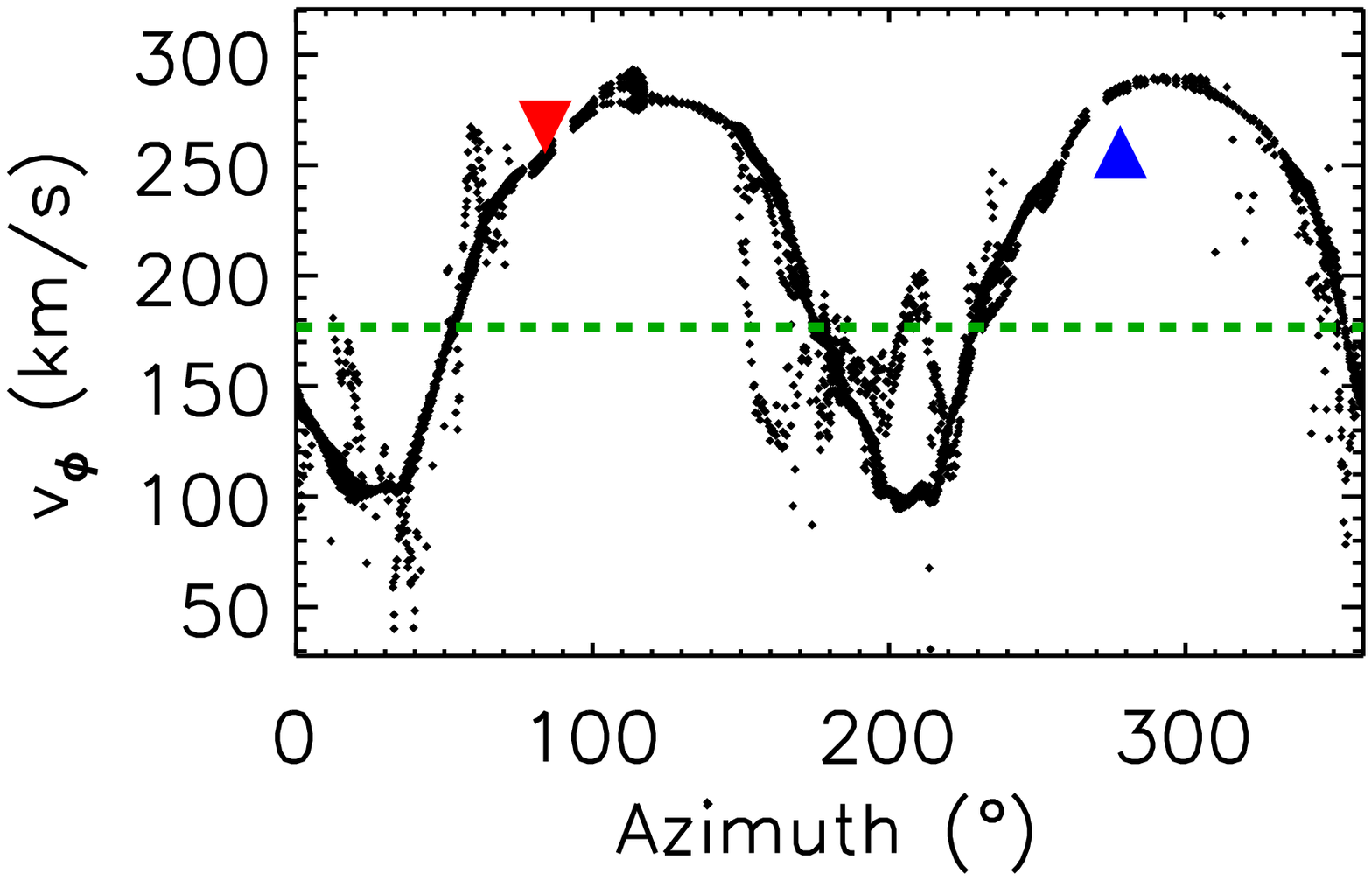}
\includegraphics[width=0.92\columnwidth]{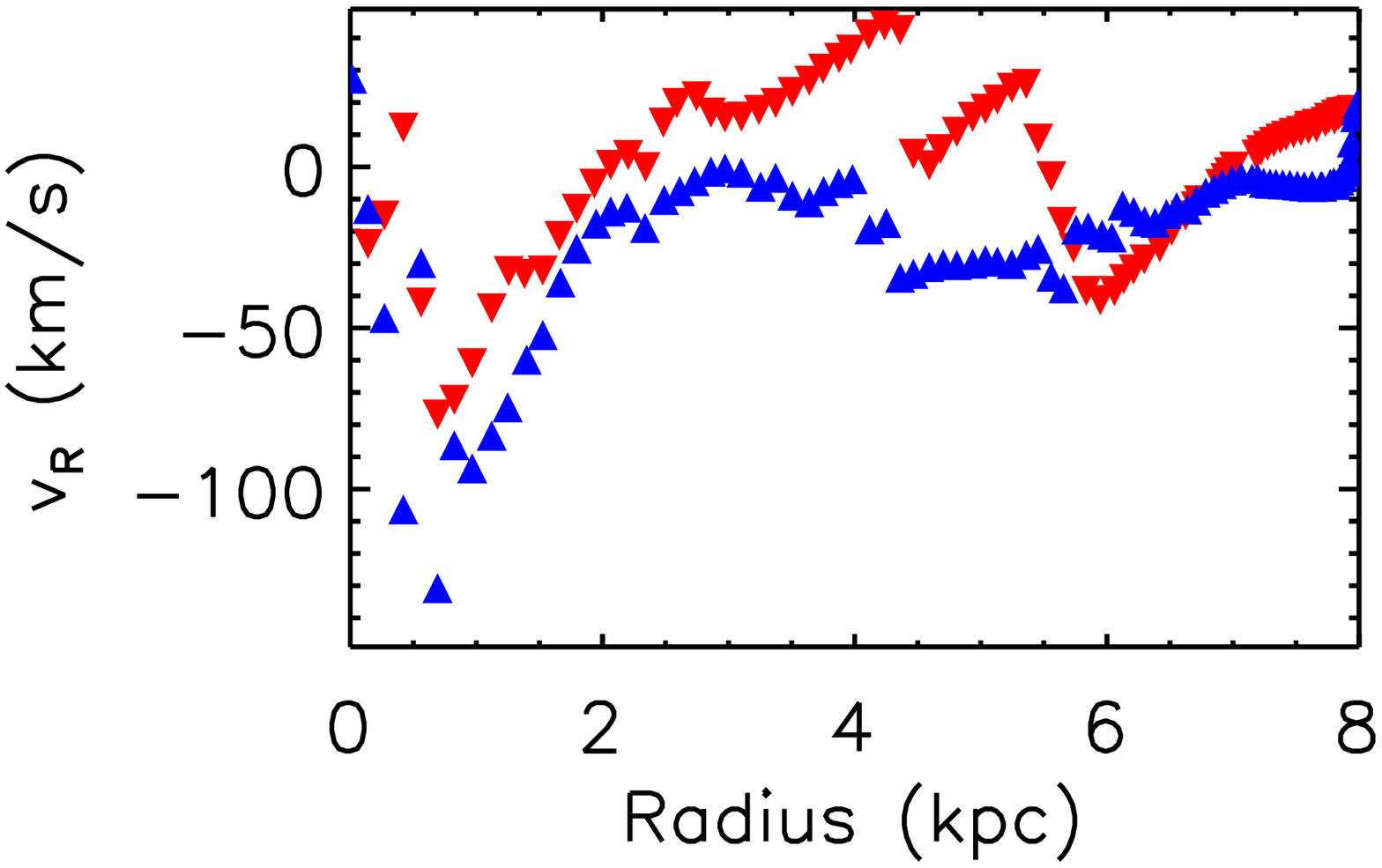}
\caption{\textbf{Top:} Azimuthal profile of tangential velocity at $R=1$ kpc. Colored symbols are the velocities inferred by the tangent-point method at a similar radius for the two quadrants. 
  The horizontal dashed line is the azimuthal  average of the velocity profile, hence the velocity of the rotation curve at $R=1$ kpc.  
  The zero azimuth corresponds to the direction of the Sun.
  \textbf{Bottom:} Profile of radial velocity $v_R$, locally given at the positions of the tangent points.  
  Each value \textup{\textit{is not}} 
  an azimuthal average from the radial velocity field at each radius. Symbols   are the same as in Fig.~\ref{fig:veltpmobs}.}
 \label{fig:compvelo1}
 \end{center}
 \end{figure}
 
The  discrepancy with the  rotation curve is  not directly linked to the tangent-point method itself. 
It is a combination of several elements: the asymmetric nature of the disk, the nonazimuthal dependency of any rotation curve,  
and the   location of the Sun. 
It is indeed  the structure of the velocity field itself  that causes the peculiar shape of the velocity profile: the tangential velocities of gas 
are highly nonuniform. Perfect axisymmetry and  uniform tangential motions  do not exist. 
This is evidenced  by the strong dependency of tangential velocities on azimuth 
 (Fig.~\ref{fig:compvelo1}, at $R=1$ kpc, e.g.).
The axisymmetric  value is $\sim$ 175 \kms, whereas the bulk of gas rotates with velocities ranging from 100 \kms\ to 290 \kms, with a few points down to 50 \kms. 
 The rotation curve smears out such nonuniform tangential motions
because a rotation curve only depends on $R$. 
   With the tangent-point method, the velocity field is unavoidably restricted  to two velocities 
  at each  radius, arising from the two quadrants. 
We showed that this is sufficient to derive the rotation curve for the early stage in the simulation since  the tangential velocity is uniform at that point, 
the disk being almost perfectly axisymmetric.
However, this is not the case anymore for the  barred and spiral configuration. 
The  location of the Sun reference point (thus the viewing angle of the bar) causes the two 
   velocities to coincide with high-velocity clouds along the bar and in   
  the spiral arms, or  low-velocity gas in lower density regions.  
  They  can therefore only be representative of \textit{local} nonuniform tangential 
motions, not of the \textit{global} rotation curve. 
Figure~\ref{fig:compvelo1} illustrates   this argument at $R=1$ kpc, where the  velocities of the two 
terminal-velocity clouds coincide with the upper envelop of the   curve.
Larger (smaller) bar viewing angles than 23\degr\ would shift the triangles to the left (right) along the curve.
At viewing angles  of $\sim$ 67\degr, the velocities  would be almost coincident with the horizontal line (azimuths of $\sim$ 50\degr\ and 230\degr), 
and the rotation curve would be correctly recovered (see also Fig.~\ref{fig:comprcveltpm}).  
 
Confirmation of the local motions affected by the bar   at low radius is also seen in the radial velocity $v_R$ profile, 
where $v_R$ are given here at the positions of the tangent-points (Fig.~\ref{fig:compvelo1}, bottom panel).
Significant noncircular motions down, for example, to $\sim -100$ \kms\ are detected exactly where
 the velocity profile of Fig.~\ref{fig:comprcveltpm} differs the most from the rotation curve, as well as at 
 larger radius. This is not fortuitous, as the asymmetries affect both the circular and noncircular motions.
 
\section{Discussion}
\label{sec:discussion}
    \subsection{Implication for the mass distribution}
\begin{figure}[t]
 \begin{center}
  \includegraphics[width=\columnwidth,trim=0 74 0 0,clip=true]{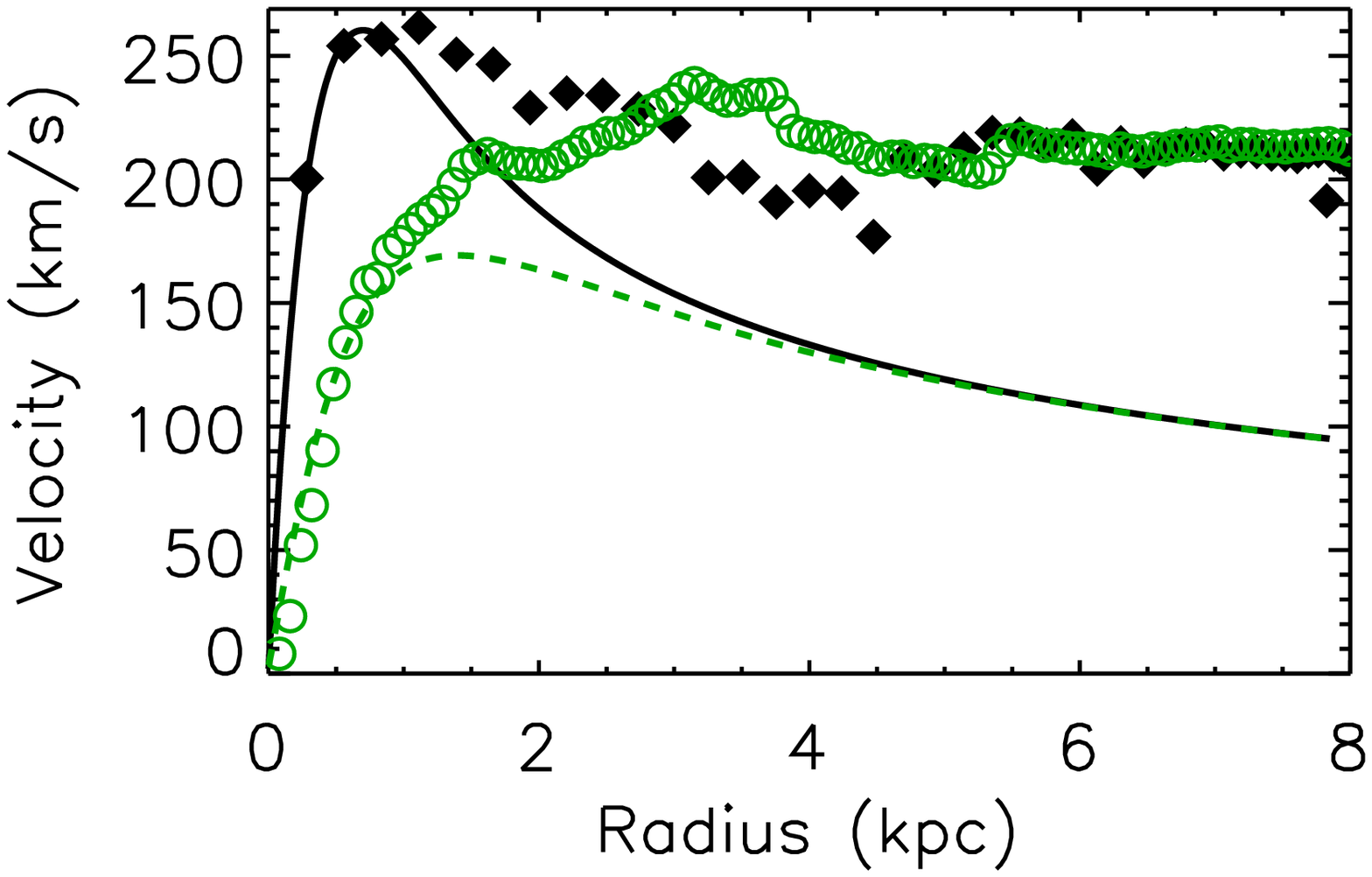}
  \includegraphics[width=\columnwidth]{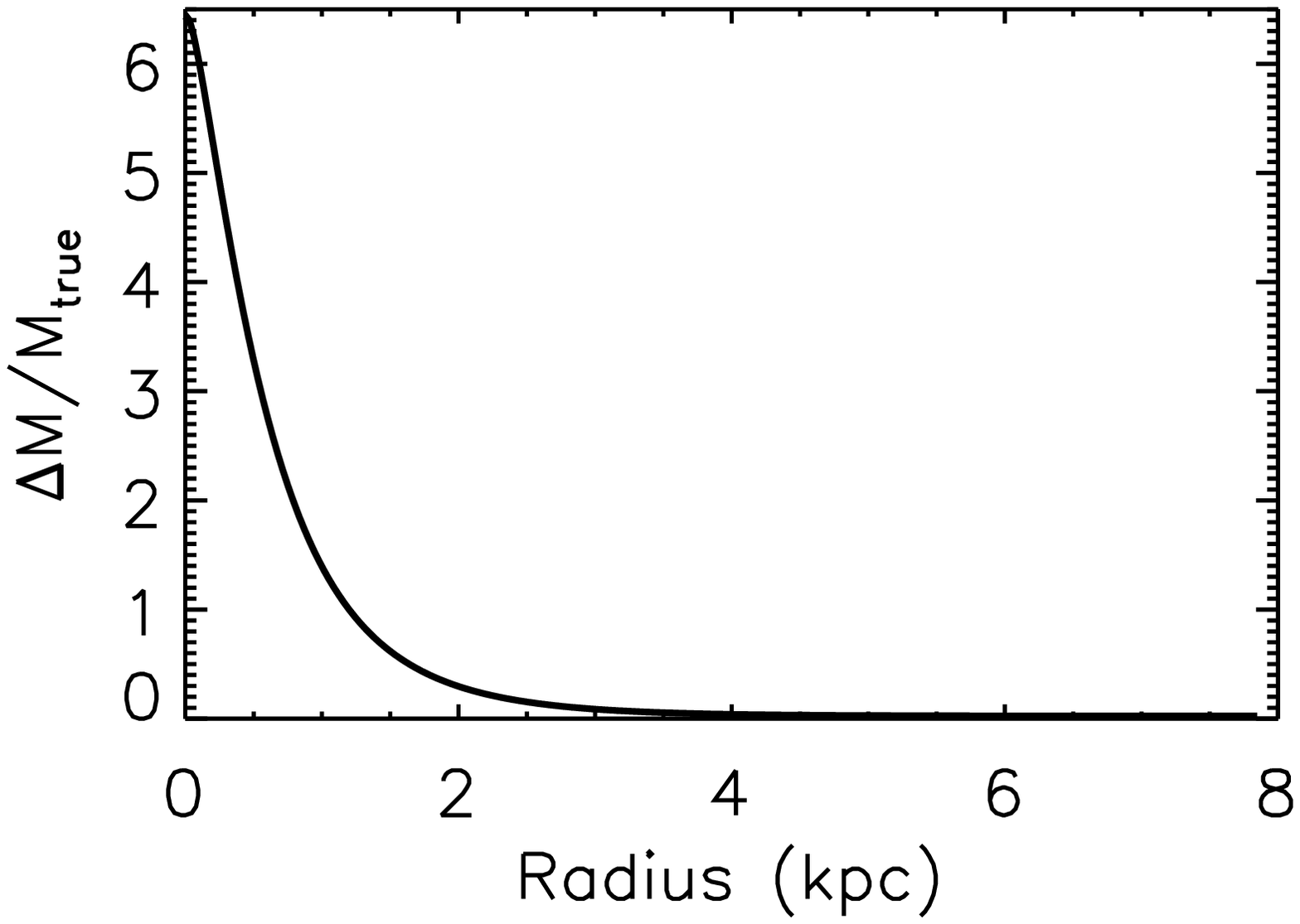}
   \caption{\textbf{Top:} Mass distribution models with the true and incorrect rotation curve of the simulated galaxy. 
   Open symbols and the green dashed line denote the true rotation curve and the bulge contribution, 
   filled symbols and a solid line represent the incorrect rotation curve and bulge contribution. 
   For clarity, the remaining contribution from   the gaseous and stellar disks plus dark matter is not shown.
   \textbf{Bottom:} Error on the integrated mass of the bulge. $\rm \Delta M=M_{inc} - M_{true}$, 
   where $\rm M_{inc}$ is the mass deduced from the incorrect  bulge, and $\rm M_{true}$ the 
   true bulge mass.}
 \label{fig:massmodel}
 \end{center}
 \end{figure}

    A straightforward implication of using the incorrect rotation curve based on the tangent-point measurements 
   is the incorrect modeling of the mass distribution at the center of the galaxy. 
   Many mass models of the Milky Way invoke a dominant contribution of the  stellar bulge   in the central regions to explain
   the steepness and the peak   of the profile seen in Fig.~\ref{fig:veltpmobs} \citep[e.g.,][]{mer92,sof09,sof13}. 
   We proceed similarly  with the simulation using the incorrect rotation curve from the fiducial case and  
   fit a bulge that best reproduces the velocities for $R < 0.8$ kpc (Fig.~\ref{fig:massmodel}). 
   
   The differences between the contributions from true bulge of the simulation   
   and the incorrect fitted bulge are shown in Fig.~\ref{fig:massmodel}.
   The true bulge contribution is not a fit to the true rotation curve, but  comes directly from the identification of stellar bulge particles 
   in the simulation.  
 
    The true and false bulges are modeled by a spherical density profile given by
         \begin{equation}
         \rho(r) =\rho_{-2} \exp\left(-2n\ ((r/r_{-2})^{1/n}-1) \right) \ .
         \label{eq:mass}
         \end{equation}
It is equivalent to an Einasto model \citep{nav04}, where $r_{-2}$ is the scale radius at which the density profile has a 
 logarithmic  slope of $-2$, $\rho_{-2}$  is the scale density at that radius, and $n$  a dimensionless index 
 describing the shape of $\rho(r)$. At fixed scale density and radius, 
 the smaller the index, the shallower the density profile in the center \citep{che11}. 

 The   parameters we  found for the incorrect  bulge   are $\rho_{-2}= 10.15\ 10^{-1}$ \msol\ pc$^{-3}$, $r_{-2}=$ 0.42 kpc, and $n=0.95$. 
The velocity peak of the incorrect bulge is 260 \kms\ at $R=0.7$ kpc, thus 54\% higher than the true bulge peak, and at a radius twice smaller. 
Note that the velocity shape and amplitude of that incorrect bulge are comparable with those seen in mass models of the Milky Way \citep{sof09}.
The Einasto  parameters of the true bulge are $\rho_{-2}= 9.46\ 10^{-1}$ \msol\ pc$^{-3}$, $r_{-2}=$ 0.87 kpc, and $n=1.14$. 
 The incorrect bulge is thus about twice  smaller and  slightly denser than the true bulge. As expected, its mass distribution 
 is therefore more concentrated than the true one for a similar enclosed masses ($\sim 1.6 \times 10^{10}$ \msol\ for the true bulge and 
$\sim 1.7 \times 10^{10}$ \msol\ for the incorrect bulge inside $R=8$ kpc).
The  effect of this higher mass concentration on the enclosed bulge mass  is shown in the bottom panel of Fig.~\ref{fig:massmodel}. 
As rule of thumb,  the considerable error on the enclosed mass  is $\sim$330\% of the true mass inside $R=0.5$ kpc, or $\sim$ 140\% inside $R=1$ kpc,
 the incorrect enclosed bulge mass being 4.3 and 2.4 times higher than the true one inside these radii.  Note that the stellar disk plus dark matter halo contribution remains 
similar in both mass models. 

 \subsection{Implications for the Milky Way}
 
In view of the results, 
the tangent-point method and terminal velocities very probably
yield an incorrect rotation curve 
of the Milky Way inside $R = 4-4.5$ kpc.  The steep central  gradient, the 
 peak at $R=0.5$ kpc, and  the smooth decrease to $R=3.5$ kpc   reflect the highly nonuniform nature of tangential velocities, 
 where gas  locally orbits faster or slower inside the Galactic bar, the spiral arms, and the interarm regions than the average at the same radii. 
 These local features cannot be representative of the true inner rotation curve of the Galaxy. 
 Beyond $R=4-4.5$ kpc ($R/R_0 > 0.5-0.56$), it is reasonable to assume that the Galactic  rotation curve  is consistent with the velocity profile obtained by the 
 tangent-point method, unless stronger   asymmetries than expected by the simulation exist at these radii in the Galactic disk.
  As a consequence,   mass distribution models of the Milky Way based on terminal velocities in the inner regions
 are unavoidably flawed.   We estimate that the Galactic bulge  
 might be less concentrated than currently thought, with a characteristic size of $\sim 1$ kpc, 
 thus about two times larger than inferred by mass models based on terminal velocities.

Furthermore, the tangent-point method gives rise to an inner steep velocity profile and a peak under particular
  circumstances, when the bar is viewed with   angles $< 45\degr$. 
  The central velocity peak is  always overestimated at viewing angles $\la 20\degr$. 
  Mock  velocity profiles with bar orientations of $23\degr$ and 30\degr\ are the models that agree best with the observational velocities. 
  Galactic bar viewing angles of $\sim 23-30\degr$ are thus favored by our analysis. 
 This agrees with many observational or numerical studies \citep[][]{rod08,weg13,ren13}, but disagrees with the observational value of $\sim$ 44\degr\ found by \cite{ham00} and \cite{ben05}. 
 In addition, the inflection  of the observed velocity curve at $R = 3.5\pm 0.3$ kpc probably marks the position of the corotation of the Galactic bar. 
 
\begin{figure}[t!]
 \begin{center}
  \includegraphics[width=\columnwidth,trim=0 0 0 0,clip=true]{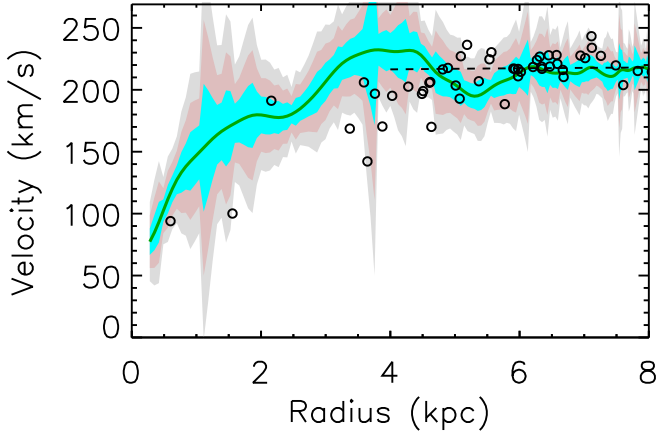}
   \caption{Hypothetical gaseous rotation curve of the Milky Way (solid line) after correcting for the initial velocity profile of Fig.~\ref{fig:veltpmobs}. 
   The cyan, pink, and gray shaded areas are the 1, 2, and 3$\sigma$ confidence levels. Open circles are 
   rotation velocities of masers from  \citet{rei14}. The dashed line is the power-law rotation curve model of the Galaxy 
   from \citet{bov12}.}
 \label{fig:compmaserapogee}
 \end{center}   
 \end{figure}   

     We can deduce a hypothetical, but more realistic, Galactic rotation curve by naively applying a correction 
  factor of $\left( \frac{v(R)}{v_{\rm true}} \right)_{\rm sim}$
to the observed profile,  which we deduce from the simulation  (see Sect. \ref{sec:results} and Fig.~\ref{fig:massmodel}). 
This first-order approach assumes that the effects inducing errors in the estimate of the rotation curve from the 
TP method in the simulations are the same as in the real Galaxy.  We have no direct way to verify the validity of this assumption  
and thus, the following hypothetical curve should be considered with caution. It gives for instance velocities that are 2.2, 1.5 
and 1.1  times lower than expected by the TP method
at  $R=0.5,1,  \text{and }  2$ kpc, respectively, and about 1.2 times higher at $R=3.5$ kpc (for $v_0=213$ \kms). 

  Figure~\ref{fig:compmaserapogee} compares 
   individual rotation velocities  from  BeSSeL, the astrometric survey of methanol and water masers in high-mass star-forming regions 
   \citep{rei09,rei14}, and the $R \ge 4$ kpc stellar 
   rotation curve from    APOGEE   \citep{bov12}, with our
    hypothetical gaseous rotation curve. Both stellar and gaseous velocities were rescaled here 
    using $v_0=218$ \kms, as given in \citet{bov12}.  
  The   velocity  uncertainty  for our gaseous rotation curve  takes into account the 
  formal error from the Gaussian fits of terminal velocities ($<$ 1 \kms), and 
    the systematic errors from (i) the mean velocity difference between profiles from
   simulated datacubes with bar orientations of 23\degr\ and 30\degr\  (5 \kms),    
    (ii)  the velocity difference $\Delta v=v(R)-v_\phi$ implied by 
  the use of TP method velocities instead of true azimuthal values  and (iii) 
  the velocity difference from both quadrants. These two latter systematic errors depend on the longitude, as seen in 
    Figs.~\ref{fig:veltpmobs} and~\ref{fig:wheretheterminalvelocitypointsare} (right panel).    
    First, it is   interesting to note that all rotation  curves are   consistent for $R>4.5$ kpc, which clearly shows that the Milky Way rotation curve is consistent with a plateau at $R > 6$ kpc.
    Secondly, our hypothetical rotation curve tends to  overestimate
    the BeSSeL velocities at $3.5 < R< 4.5$ kpc, though remaining in agreement within the 3$\sigma$ confidence level. 
    This is probably due to  systematic effects inherent to the construction of our hypothetical curve, that is, to
    the TP method  and the  simulation themselves.   Note, however, that depending on the location of the masers
     down- or up-stream of the peak densities inside the spiral arms,
     velocities from BeSSeL may not coincide perfectly with the mean rotation velocities  either. They might trace 
     faster or slower \textit{\textup{local}} motions, exactly  like those seen for individual gas clouds
      selected by the TP method (Fig~\ref{fig:compvelo1}; see also Fig. 4 in \citeauthor{kaw14} \citeyear{kaw14}). 
    It appears  essential here  that the inner Galactic velocity field has to be covered as uniformly as possible 
    in radius and azimuthal angle to obtain the most accurate inner rotation curve,   similarly as is routinely 
       done for nearby galaxies with 3D  spectroscopy or radio interferometry \citep[][]{ems04,che06,deb08,gar15}.
    This would require observations of  other tracers than masers in more diffuse regions  outside the bar  and the arms. 
    This is not yet possible for VLBI, however.
     
 A final consequence of the analysis is that rotation curves of edge-on galaxies  based on envelope-tracing of 
 \hi\ terminal velocities \citep[e.g.,][]{swa97,kre04} 
  might   be systematically biased toward too high or low velocities as well. Unless no bar or  strong perturbation 
  is shown to exist in their central regions, caution should be taken when modeling these velocity curves because  
 a steep (shallow, respectively) velocity rise could mimic a stellar  component that is too (less) concentrated, 
 or a  dark matter halo that is too cuspy \citep[shallow; see also][for less inclined disk cases]{rhe04,val07,dic08}.

\begin{acknowledgements}
 We are grateful to Fr\'ed\'eric Bournaud and Justin Read for insightful discussions. 
 We also thank the anonymous referee for constructive comments and suggestions. 
 LC acknowledges financial support from Centre National d'\'Etudes Spatiales.  
FR acknowledges support from the European Research Council through the grant ERC-StG-335936.
\end{acknowledgements}

\bibliographystyle{aa}
\bibliography{chemintpm}

\end{document}